\documentclass[superscriptaddress,twocolumn,aps,prb,floatsfix,showpacs]{revtex4}
\usepackage{graphicx}
\usepackage{amssymb,amsmath}
\usepackage{array}
\usepackage{dcolumn}
\usepackage[colorlinks,bookmarks=false,citecolor=blue,linkcolor=blue,urlcolor=blue]{hyperref}
\usepackage{color}

\begin{document}
 
\title{A $2D$ quantum antiferromagnet with a  four-fold 
degenerate valence-bond-solid ground state}

\author{Alain Gell\'e} \affiliation{Institute of Theoretical Physics, \'Ecole
  Polytechnique F\'ed\'erale de Lausanne, CH 1015 Lausanne, Switzerland}

\author{Andreas M. L\"auchli} \affiliation{Institut Romand de Recherche
  Num\'erique en Physique des Mat\'eriaux (IRRMA), CH-1015 Lausanne, Switzerland}

\author{Brijesh Kumar} \affiliation{School of Physical Sciences, Jawaharlal
  Nehru University, New Delhi 110067, India}

\author{Fr\'ed\'eric Mila} \affiliation{Institute of Theoretical Physics, \'Ecole
  Polytechnique F\'ed\'erale de Lausanne, CH 1015 Lausanne, Switzerland}

\date{\today}
 
\begin{abstract}
  
  We study the competition between antiferromagnetic order and valence bond
  solid formation in a two-dimensional frustrated spin-1/2 model. The
  $J_1$-$J_2$ model on the square lattice is further frustrated by introducing
  products of three-spin projectors which stabilize four dimer-product states
  as degenerate ground states. These four states are reminiscent of the
  dimerized singlet ground state of the Shastry-Sutherland model. Using exact
  diagonalizations, we study the evolution of the ground state by varying the
  ratio of interactions.  For a large range of parameters ($J_2 \gtrsim
  0.25J_1$), our model shows a direct transition between the
  valence-bond-solid phase and the collinear antiferromagnetic phase. For
  small values of $J_2$, several intermediate phases appear which are also
  analyzed.

\end{abstract}
\pacs{75.10.Jm, 75.30.Kz, 75.50.Ee, 75.40.Mg }
                                                                                
\maketitle
                                                                                
\section{Introduction}

In the last several years, a large amount of work has been devoted to the
study of quantum systems with frustrated magnetic interactions due to their
propensity to present spin liquid states. These phases indeed present
interesting low energy properties, starting with the absence of magnetic order
even at zero temperature. Recently, a lot of attention has been focused
particularly on the Valence Bond Solids (VBS), and more precisely on the
transition between these states and a magnetically ordered phase. In this
context, it has been suggested that this transition could be a second order
transition that does not belong to the Landau-Ginsburg
paradigm~\cite{DQCP_1,DQCP_2}.

In a VBS state, spins are coupled in pairs forming singlet states, evocating
valence bonds. These pairwise singlets are themselves arranged in a periodic
pattern. Such states break the translation symmetry, and it is possible to
define an order parameter quantifying the singlet long range order. The
question on the nature of quantum phase transition that separates this state
from a long-range ordered magnetic phase is very interesting as well as
current. Supposing a second order transition, according to Landau-Wilson
paradigm~\cite{Landau}, the order parameter of both phases should vanish
precisely at the transition. It seems more likely that the two parameters will
not vanish exactly at the same point, leading either to a first order
transition, or to two second order transitions separated by an intermediate
phase. Considering spin-1/2 on square lattice, Senthil and coworkers have
recently suggested that there could exist second order transitions
that are not described by Landau-Ginsburg theory~\cite{DQCP_1,DQCP_2}. The
transition could instead be described by means of fractional degrees
of freedom, namely spinons. These spinons become deconfined at the transition
point, called the deconfined quantum critical point.

A priori, the spin-1/2 $J_1$-$J_2$ antiferromagnetic model on the square
lattice seems the simplest choice to investigate the relevance of this theory.
In the two limiting cases where either $J_1$ or $J_2$ is very large compared
to the other, the system presents antiferromagnetic order. In the former case
it corresponds to the usual N\'eel order, and in the latter case the ground
state has a collinear antiferromagnetic order, corresponding to the N\'eel
order on two sublattices (that are obtained by connecting second
neighbor sites). It is generally accepted that this model
presents an intermediate spin disordered phase in the range
$0.4<J_2/J_1<0.6$, which might break the translational
symmetry~\cite{JJ_ED96,JJ_MC01,JJ_ED91,JJ_Series00,JJ_Series99,JJ_Pert96,JJ_MC00}.
It is still unclear if the transition between the N\'eel state and the spin
disordered phase is a deconfined quantum critical point or a simple
first order transition, although recent works are in favor of a weak first
order transition~\cite{JJ_MC04,JJ_Series06}.

The difficulty in understanding the $J_1$-$J_2$ model comes from the
intermediate phase whose nature is subject to
discussions~\cite{J3_ED06}. Some studies show a four-spin plaquette
order~\cite{JJ_Pert96,JJ_MC00}, while others are in favor of a columnar dimer
order~\cite{JJ_ED91,JJ_Series00,JJ_Series99}. In the latter case, not only the
translational symmetry, but also the rotational symmetry is broken. 
However, if this phase develops dimer-dimer correlations, it is far away from
being the simple direct product of dimer singlet wave functions. Indeed, if that
was the case,
the ground state would present a strong signal of long range order
which is not observed for the intermediate phase of the $J_1$-$J_2$ model.
Exact diagonalization studies show that the dimer-dimer correlation is rapidly
decreasing with distance~\cite{JJ_ED91,JJ_ED03}, and that, if the dimer-dimer
order exists, it should be rather small~\cite{JJ_ED03}. The analysis of the
properties of the phase transition is therefore not so easy in the $J_1$-$J_2$
model. Recently, a new model including ring exchange has been
proposed~\cite{ProjPlaq} to explore the possibility of non Landau-Ginsburg
phase transitions. It has however been shown that this model presents a first
order transition~\cite{ProjPlaq}.

In the present paper, we propose a new quantum spin-1/2 model with
frustrated antiferromagnetic interactions. Interestingly, for a simple choice
of the interaction parameters, this model has an exactly solvable singlet
ground state. The exact ground state is a pure direct product of dimers,
arranged in the same pattern as that in the ground state of the
Shastry-Sutherland model~\cite{SS}. However, unlike in the Shastry-Sutherland
model, the Valence Bond Solid (VBS) ground state
of our model presents a case of spontaneous symmetry breaking, and is
four-fold degenerate~\cite{Read_Sachdev,Haldane}. Using exact diagonalization,
we investigate the evolution of the ground state as the interaction parameters
are varied away from the exactly solvable VBS case, particularly towards
magnetically ordered phases.
In the following section, we describe the model and discuss its main 
properties and its relation to the $J_1$-$J_2$ model. In section III-A, we 
study the competition between the VBS phase and the antiferromagnetic 
collinear phase. In section III-B, we analyze the competition with the usual 
N\'eel antiferromagnetic phase. Finally, the last section is devoted to 
conclusions and perspectives.

\section{model}

\begin{figure}[htbp] 
\resizebox{8cm}{!}{\includegraphics{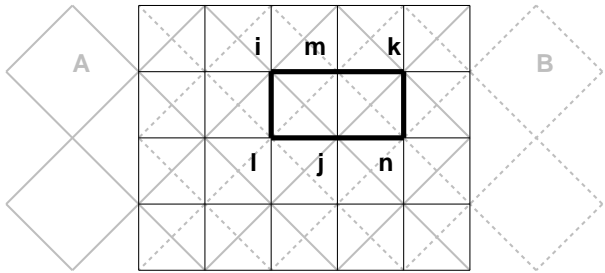}}
\caption{Representation of a $3$ by $2$ horizontal rectangular plaquette. A vertical $2$ by
  $3$ plaquette can be obtained by a $90^\circ$ rotation of the plaquette.
  Gray plain and dashed lines represent the nearest-neighor bonds of the A and
  B sub-lattices.}
\label{f:plaq}
\end{figure}

Consider a system of spin-1/2 on the square lattice. The model we consider
contains only two and four spins interactions which can be conveniently
described using a six-site rectangular plaquette (see Fig.~\ref{f:plaq}). It is also
convenient to distinguish between the two sublattices A and B of the square
lattice. The six sites of a plaquette thus contain 3 sites belonging to the
$A$ sublattice and other $3$ sites belonging to the $B$ sublattice. For each
set of three spins, we consider the spin projector on the quartet (spin 3/2) state:
\begin{eqnarray}
  P_{i,j,k}^{A} &=& |\vec S_i+\vec S_j+\vec S_k|^2 - \frac{3}{4}\\
  P_{l,m,n}^{B} &=& |\vec S_l+\vec S_m+\vec S_n|^2 - \frac{3}{4}
\end{eqnarray}
where A and B refer to the two sublattices, and $i,j,k,l,m,n$ are the sites of
the plaquettes as depicted in Fig~\ref{f:plaq}. The interaction we consider is
obtained by taking the product of the two projectors of a plaquette:
\begin{equation}
  H_0= \sum_{[i,j,k,l,m,n]} \frac{1}{4}\; P_{i,j,k}^{A}P_{l,m,n}^{B}
\end{equation}
where the sum runs over all horizontal and vertical plaquettes, and
$[i,j,k,l,m,n]$ are the sites of one plaquette.

One should notice that the value of a projector is always positive except when
the three spins are in a doublet state (that is when it becomes zero).  The
latter condition is fulfilled, in particular, if any two of the three spins
form a singlet. These two spins could either be two first neighbor or two
second neighbor spins of a sub-lattice, and correspond respectively to two
second or two third neighbor spins of the original square lattice. Since the
interaction in the model is the product of two projectors, it is only
necessary to cancel one of these projectors to minimize the corresponding
term. For this purpose, it is thus sufficient to have one dimer on the
plaquette, either between second or third neighbor spins.  Therefore, it is
obvious that if a valence bond configuration has one such dimer on every
plaquette, it will form a zero energy ground state of $H_0$.

\begin{figure}[htbp] 
\resizebox{4.44cm}{!}{\includegraphics{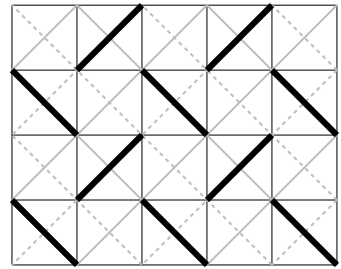}}
\caption{Representation of one of the four ground states of model $H_0$.
  Black lines indicate the pairs of spins that form singlet states. The
  three other ground states can be obtained by translation of the dimer
  pattern by lattice vectors.}
\label{f:ss}
\end{figure}
The idea of stabilizing valence bond states by spin projectors is also
contained in the Majumdar-Ghosh model on the linear chain~\cite{MajGos}.
Recently, Batista and Trugman studied a model with such projectors on
four-site square plaquettes~\cite{ProjPlaq}, but their Hamiltonian has a
highly degenerate ground state since a great number of dimer patterns minimize
all projectors.  In the present model, the ground state is obtained when
dimers are arranged in the pattern given in Fig.~\ref{f:ss} and is much
  less degenerate (see also the discussion in the Appendix). This arrangement evokes the
Shastry-Sutherland model (and we will further refer to this phase as the
SS-VBS phase), but contrary to the SS model, our Hamiltonian does not break
the translational symmetry of the square lattice. The ground state presents a
spontaneous symmetry breaking, and is four-fold degenerate.  The other three
of SS-VBS states can be obtained by translation of the dimer pattern by
lattice vectors. Interestingly, the four SS-VBS states are also exact zero
energy eigenstates (although not the ground state) of the nearest-neighbor
Heisenberg model on the square lattice, which motivated us to seek for a model
on the square lattice with four SS-VBS states as the exact ground
state~\cite{BK}.

The spin projectors are two-spin operators which can be rewritten in terms of the 
exchange couplings between the first and second neighbor spins of a sublattice:
\begin{equation}
  P_{i,j,k}^{A} = 2(\vec S_i\cdot\vec S_j+\vec S_i\cdot\vec S_k+\vec S_j\cdot\vec S_k) + \frac{3}{2}
\end{equation}
Therefore, the product of two projectors generates two- and four-spin
operators in $H_0$. As the projector is invariant under rotation and involves spins
on a given sublattice, it will commute with the total spin of this sublattice,
and obviously, with the total spin of the other sublattice. $H_0$ will
thus conserve the total spin on each sublattice.  This extra symmetry may
introduce a further factor two in the degeneracy of excited states.

It is interesting to notice that this Hamiltonian is somehow related to the
$J_1$-$J_2$ model. Let us assume that a state can be expressed as a direct
product of the wavefunction of spins on A sublattice $|\Psi_A \rangle$ and of
spins on B sublattice $|\Psi_B \rangle$~:
\begin{equation}
|\Psi \rangle = |\Psi_A \rangle \otimes |\Psi_B \rangle
\end{equation}
In that case, it is possible to define a Hamiltonian for one sublattice. Let us, for 
instance, choose the A sublattice:
\begin{equation}
H_A =  \sum_{[i,j,k,l,m,n]} \frac{1}{4}\; P_{i,j,k}^{A} \langle \Psi_B | P_{l,m,n}^{B}| \Psi_B \rangle
\end{equation}
This Hamiltonian corresponds, for the A sublattice, to a model with first and
second neighbor interactions whose amplitudes are modulated by the local spin
state of the B sub-lattice.  It is easy to show that, in the case where the
mean value of the projector on B sublattice is homogeneous, the Hamiltonian
can be expressed as:
\begin{equation}
H_A = \langle P^{B} \rangle \sum_{<i,j>_A} \vec S_i\cdot\vec S_j
+\frac{1}{2} \langle  P^{B} \rangle \sum_{\ll i,j\gg_A } \vec S_i\cdot\vec S_j
\end{equation}
where $<i,j>_A$ are the first neighbor couples of spins of $A$ sublattice, and
$\ll i,j\gg_A $ the second neighbor ones, $\langle P^{B} \rangle$ is the
uniform value of $\langle \Psi_B | P_{l,m,n}^{B}| \Psi_B \rangle$. This
Hamiltonian is precisely the $J_1$-$J_2$ model on A sublattice at the
$J_2/J_1=0.5$ point, which is located in the intermediate spin liquid phase of
the model. Clearly, if we consider the ground state of our model, we do not
expect the mean value of the $P^B$ projector to be homogeneous. Nevertheless,
this comparison introduces a simple picture of the model, in which the local
spin correlations self-consistently modulate the $J_1$-$J_2$ model on each
sub-lattice. This process allows the system to stabilize a pure direct product
of dimers as ground state. This valence bond ground state is
different in nature from the one expected in the intermediate phase of
$J_1$-$J_2$ model, since the dimer order on each sub-lattice is not columnar but
staggered~\cite{RingExchangeSquare} (see Fig.~\ref{f:ss}).

\begin{figure}[htbp] 
\includegraphics[width=0.8\linewidth]{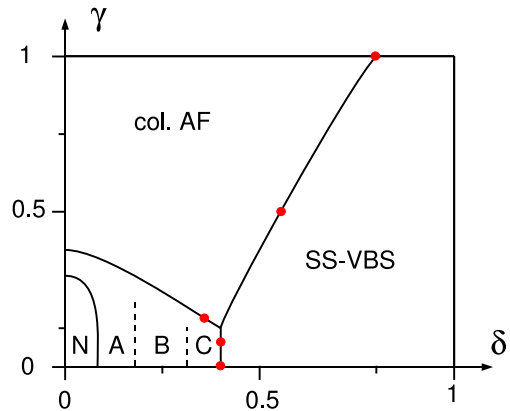}
\caption{(color online) Sketch of the phase diagram of the model
  $(\protect{\ref{eqn:Hamiltonian}})$.  Phase transitions obtained by exact
  diagonalizations are indicated by full dots. The lines are guide to the
  eyes. The nature of the different phases is described in the following
  sections.
\label{fig:phase_diagram}
}
\end{figure}
The existence of this exact ground state makes this model an interesting
candidate to explore the transition between the VBS phase and the
antiferromagnetically ordered state. In order to drive the system into such
phases, we consider the following Hamiltonian:
\begin{eqnarray}
H&=& J(1-\gamma)(1-\delta) \sum_{<i,j>} \vec S_i\cdot \vec S_j \nonumber \\ 
&&+J\;\gamma(1-\delta) \sum_{\ll i,j\gg} \vec S_i\cdot \vec S_j + J\;\delta\; H_0
\label{eqn:Hamiltonian}
\end{eqnarray}
where $<i,j>$ and $\ll i,j\gg$ are respectively first and second neighbor
spins of the original (square) lattice, $\gamma$ and $\delta$ are dimensionless
parameters, and $J$ is the energy scale (assumed to be positive). 
When $\delta$ is equal to $1$, one just
retrieves the $H_0$ model, while  when $\delta=0$ is zero, 
the Hamiltonian is simply the $J_1$-$J_2$ model, with
$\gamma$ being equal to $J_2/(J_1+J_2)$. A sketch of the phase diagram of
the Hamiltonian (\ref{eqn:Hamiltonian}) is shown in Fig.~\ref{fig:phase_diagram}.
It contains three phases where the order has been unambiguously identified
denoted by N (antiferromagnetic $(\pi,\pi)$ N\'eel order), col. AF 
(collinear $(\pi,0)/(0,\pi)$ order with N\'eel order on the two sublattices) and SS-VBS (Valence-Bond
order with Shastry-Sutherland arrangement). Between these phases, there is 
a region where correlations change very significantly, defining possible phase
transitions between three phases denoted by A, B and C. These phases are
discussed in the last part of the next section. 

\section{Numerical Results}

We study the evolution of eigenstates of this model by an exact
diagonalization technique based on the Lanczos algorithm. Finite clusters of
$N=16$, $20$ and $32$ sites were used. Larger clusters were not accessible
because of the complexity of the interaction.

\begin{figure}[htbp] 
\resizebox{3.5cm}{!}{\includegraphics{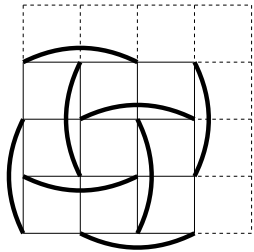}}
\caption{Representation of one of the two additional ground states for the $16$ site
  cluster. Full lines connect the $16$ sites of the cluster, dashed lines
  symbolize the boundary conditions. The other additional ground state can be
  obtained by translation of the dimer pattern by lattice vectors.
}
\label{f:ss16}
\end{figure}
The first noticeable result concerns the $\delta=1$ case, when $H=JH_0$.  We
checked numerically that the four SS ground states presented in previous
section are the only ground states of the model. This is always true except
for the $16$ site cluster, for which two more ground states are also present,
as shown in Fig.~\ref{f:ss16}.  These two states are ground states only for
the $16$ site cluster, due to the very short loops which wrap around the
boundaries of the cluster (see also the Appendix).
Indeed, in this case a third neighbor dimer between the sites $(x,y)$ and 
$(x,y+2)$ also represents a dimer between the sites $(x,y+2)$ and $(x,y+4)$
because of the periodic boundary conditions. This is no longer true for
larger clusters and by extension in the thermodynamic limit.

In the following two subsections we map out the phase diagram of the model,
with emphasis on two lines:
 In subsection~\ref{s:CO} we study first the case
$\gamma=1$, which connects to the case of $J_1=0$, $J_2=J$ once $\delta=0$. 
In subsection~\ref{s:NE} the case $\gamma=0$ is presented,
which connects the unfrustrated square lattice Heisenberg antiferromagnet to
the fourfold dimerized ground state.

\subsection{SS-VBS versus Collinear Order ($\gamma = 1$)}
\label{s:CO}

In the present section, we consider the case where $\gamma=1$.  According to
the previous discussion on the relation between our model and the $J_1$-$J_2$
model, $J_2$ corresponds to a first neighbor interaction on each sublattice
which, for large enough values, will lead to a N\'eel state on each
sublattice.

\begin{figure}[htbp] 
  \resizebox{7cm}{!}{\includegraphics{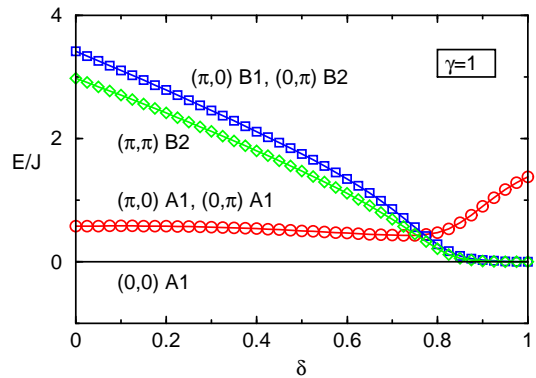}}
\caption{(color online) Energy differences between the ground state and some of the lower
  states of the $32$ site cluster. The symmetry of the different states is
  indicated on the figure. States with $(\pi,0)-A_1$ and $(0,\pi)-A_1$
  symmetry are triplet states, others are singlet states.}
\label{f:j2e32}
\end{figure}
For $\delta=1$, the SS-VBS states are exact ground states even for finite
clusters. Note that these four states are nonorthogonal on a finite cluster:
they have a finite overlap that decreases exponentially with the cluster
size~\cite{DOverlap}.  However,
since they are linearly independent, the ground state is indeed four-fold
degenerate on a finite cluster for $\delta=1$.  This degeneracy is lifted
however by the $J_2$ interactions that appear for $\delta<1$.
Fig.~\ref{f:j2e32} represents the energy differences between the ground state
and the lowest state of some symmetry sectors, obtained for the $32$ site
cluster. The energy differences between the four lowest singlet states stay
relatively small in the range $0.8<\delta\leq 1$, while the spin gap
progressively decreases. Below $0.8$ the energy difference between singlet
states rapidly increases, while the spin gap remains approximatively constant
suggesting that the system is in the antiferromagnetic phase that will present
a zero spin gap in the thermodynamic limit.

\begin{figure}[htbp] 
\resizebox{7cm}{!}{\includegraphics{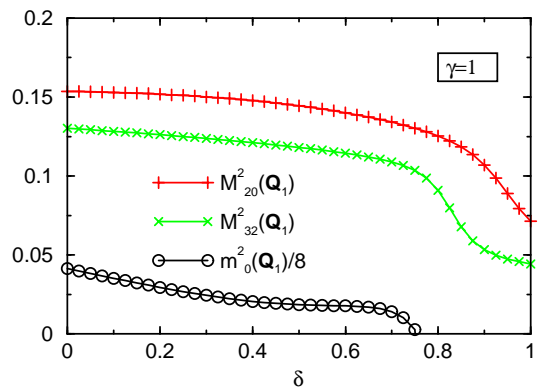}}
\caption{(color online) $M_N(\vec Q_1)^2$ square magnetic susceptibility for $20$ and $32$ sites
  clusters and its extrapolation to thermodynamic limit $M_\infty(\vec
  Q_1)^2=m_0^2(\vec Q_1)/8$.}
\label{f:m0P}
\end{figure}
In order to further investigate the transition, we calculate the  
$\vec Q$ dependent magnetic susceptibility for each cluster~\cite{kiNN2}:
\begin{equation}
M_N^2(\vec Q)=\frac{1}{N(N+2)} \sum_{i,j} \langle\vec S_i \vec S_j\rangle e^{i\vec Q(\vec
  r_j-\vec r_i)}
\end{equation}
where $\vec{r}_i$ denotes the position of $i^{th}$ spin, $<>$ the mean value 
in the ground state and $N$ the number of sites in the cluster. The evolution
of the $\vec{Q}_1=(\pi,0)$ magnetic susceptibility relevant for antiferromagnetic
collinear order is presented in Fig.~\ref{f:m0P}.  The extrapolation to the
thermodynamic limit of the corresponding sublattice
magnetization~\cite{JJ_ED96} has been performed using the finite size scaling
predicted by non-linear sigma model studies~\cite{FSE89,FSE93}:
\begin{equation}
M_N^2(\vec Q_1)= \frac{1}{8} m_0^2(\vec Q_1) + \frac{\rm const.}{\sqrt{N}}
\end{equation}
The fitted value of $m_0(\vec Q_1)$ is also shown in Fig.~\ref{f:m0P}. The
extrapolated magnetization stays large up to $\delta\simeq 0.7$, which
confirms that the collinear phase is stable in this range of parameter. It
then rapidly drops, and vanishes around $\delta\approx0.75$.

\begin{figure}[htbp] 
\resizebox{7cm}{!}{\includegraphics{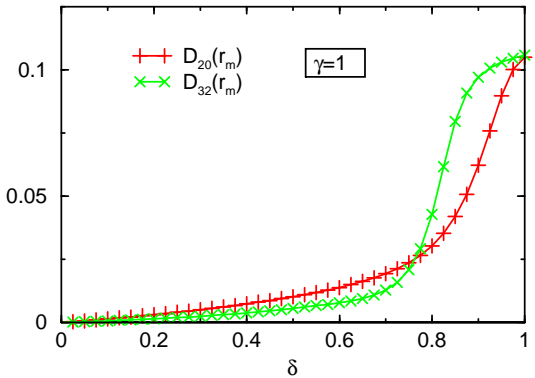}}
\caption{(color online) Dimer-dimer correlation $D_m$ obtained for the $20$ and $32$ sites
  cluster. This value correspond for one cluster to the correlation between
  the two dimers of the Shastry-Sutherland pattern separated by the largest
  distance.}
\label{f:dd}
\end{figure}
The VBS phase, which is expected for larger values of $\delta$, is
characterized by long range dimer-dimer correlation. This long range
correlation can be considered as the order parameter of this symmetry-breaking
phase. In order to determine the stability of the phase, we compute the
following dimer-dimer correlations in the ground state:
\begin{equation}
D_N(\vec r)= \langle(\vec S_{\vec 0}\cdot \vec S_{\vec r_1}) (\vec S_{\vec r}\cdot \vec
S_{\vec r + \vec r_2})\rangle - \langle\vec S_{\vec 0}\vec S_{\vec r_1}\rangle\langle\vec S_{\vec
  0}\vec S_{\vec r_2}\rangle
  \label{eqn:dimercorrs}
\end{equation}
where $\vec 0$ stands for the origin, and $\vec r_1$ and $\vec r_2$ can be
either equal to $(1,1)$, $(1,-1)$, $(-1,1)$ or to $(-1,-1)$. As expected, for
$\delta\sim 1$, values of $D(\vec r)$ that corresponds to the dimers of the
Shastry-Sutherland pattern are quite large. For a given cluster, and close to
$\delta=1$, fluctuations of these values are very small, of the order of a few
percent of the average value of these correlations. Since we are interested in
the value of $D(\vec r)$ for $r$ going to infinity, we only considered among
these correlations the one obtained for the largest $r$ value ($D_N(\vec
r_m)$). These correlations, shown in Fig.~\ref{f:dd}, are quite small in the
antiferromagnetic phase, and rapidly increase at around $\delta\sim 0.8$.
Interestingly enough, the curves cross at $\delta=0.76$, very close to 
the point were the antiferromagnetic order vanishes. This behaviour
is consistent with a first order transition, with an order parameter
scaling down to zero with the cluster size below a critical 
value, and scaling up to a finite value above. However, with only two
sizes available (the 16 site cluster turns out to be rather pathological
with essentially $\delta$ independent correlations), this information
should be taken with care, and a definitive identification of the nature
of the phase transition requires further investigation.

\subsection{Competition with N\'eel order ($\gamma=0$)}
\label{s:NE}
We now consider the case of $\gamma=0$, which corresponds to the competition
between the first neighbor coupling and the six sites plaquette interaction.
The interest in the model obtained for $\gamma=0$ comes from the fact that the
SS-VBS states remain eigenstates for all values of $\delta$ with an
energy equal to zero. It follows that the transition at which the valence bond
state vanishes is necessarily a level crossing and is therefore first order.

It should be emphasized here that, in the previous section, both SS-VBS and
the collinear order phases presented antiferromagnetic correlations between
second neighbor spins. Therefore, a direct transition between them seems to
intuitively exist as if the spin-liquid phase (SS-VBS) knows, through the
spin-spin correlations present in it, to which ordered phase it must go. This
is no more the case for the N\'eel order which is stabilized by the first
neighbor coupling, and presents antiferromagnetic correlations between
first neighbor spins, and ferromagnetic correlations between second
neighbor spins. According to this simple consideration, a direct transition
between the SS-VBS phase and the N\'eel phase seems therefore
unlikely. 

\begin{figure}[htbp] 
  \resizebox{8cm}{!}{\includegraphics{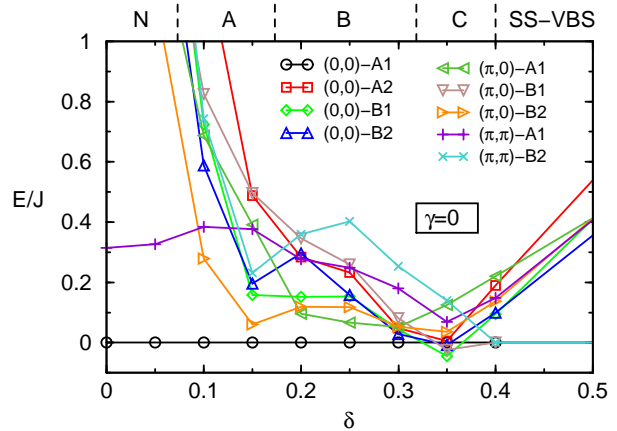}}
\caption{(color online) Energy differences between the lower fully symmetric state with
  $\vec k=(0,0)$ momenta and some of the lowest states of the $32$ site
  cluster. The symmetry of the different states is indicated on the figure.}
\label{f:j1e32}
\end{figure}

\begin{figure}[htbp] 
  \resizebox{7cm}{!}{\includegraphics{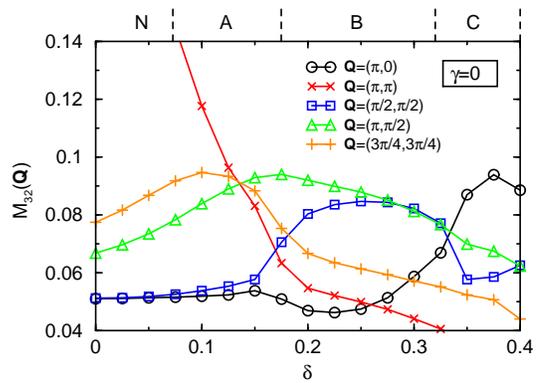}}
\caption{(color online) $Q$ dependent magnetic susceptibility obtained for the
  $32$ sites cluster.}
\label{f:J1mQ}
\end{figure}

We start the discussion by presenting in Fig.~\ref{f:j1e32} the evolution of the energies 
of some of the lowest eigenstates obtained for the $32$-site cluster, taking the lowest 
fully symmetric $\vec k=(0,0)$ level as the energy reference. Based on this figure and 
Fig.~\ref{f:J1mQ}, we are lead to identify three different phases, tentatively labeled
"A", "B" and "C" in addition to the well characterized $(\pi,\pi)$ N\'eel phase at $\delta=0$ and the
fourfold degenerate Shastry-Sutherland states for $\delta \gtrsim 0.4$. We discuss each
of these three phases in the following.

\begin{figure}[htbp] 
  \resizebox{!}{6cm}{\includegraphics{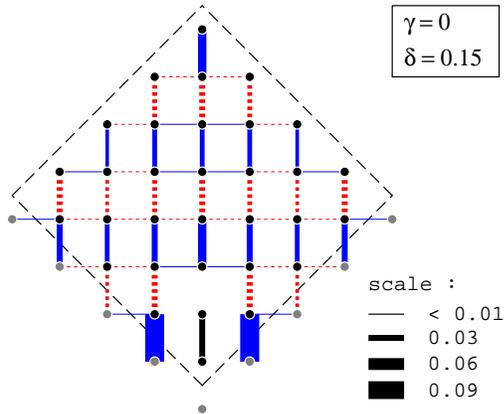}}
\caption{(color online) Correlations between first neighbor spin dimers on
  the $32$ sites cluster for $\gamma=0$ and $\delta=0.15$. Positive values are
  represented by (blue) plain lines, and negative values by (red) dashed
  lines. The thickness is proportional to the amplitude of the correlation as
  depicted on the figure. The (black) dimer in the lowest corner of the cluster is the
  reference dimer.}
\label{f:CDim}
\end{figure}
{\em Phase "A" -- } Starting at small values of $\delta$, we note that the plaquette interactions
favor antiferromagnetic correlations between second neighbor spins. Therefore
one could expect the evolution of the system to be similar to the $J_1$-$J_2$
model for small values of $J_2$. The symmetry of the four lowest states near
$\delta=0.15$ with momenta $(0,0)$ (two states), $(\pi,0)$ and $(0,\pi)$, are
indeed compatible with the hypothesis of a four-fold degenerate ground state
with a translational symmetry breaking. Fig.~\ref{f:CDim} shows the
correlations defined in Eqn.~(\ref{eqn:dimercorrs}), this time between dimers made 
of first neighbor spins obtained for $\delta=0.15$. These correlations are relatively 
large, and show a columnar ordering of the dimers. Nevertheless, one should note 
that, as for the $J_1$-$J_2$ model, it is difficult to determine whether this
phase presents columnar dimer order or plaquette order in the thermodynamic limit. 
So we believe the phase "A" is formed of some sort of valence bond solid with dimers
on nearest-neighbor sites.

{\em Phase "C" -- }
At the other end of the $\delta$ axis, starting from large values of $\delta$, 
one can see that the ground state stays exactly fourfold degenerate down to 
$\delta\simeq 0.4$, below which a level crossing occurs. Near $\delta=0.35$ 
many levels are very close in energy. 
Some of these states are even found to be lower in energy than the fully symmetric
$\vec k=(0,0)$ state -- which is again the ground state for $\delta \le 0.3$, 
but this may well be a finite size effect on this particular sample.

\begin{figure}[htbp] 
  \resizebox{7cm}{!}{\includegraphics{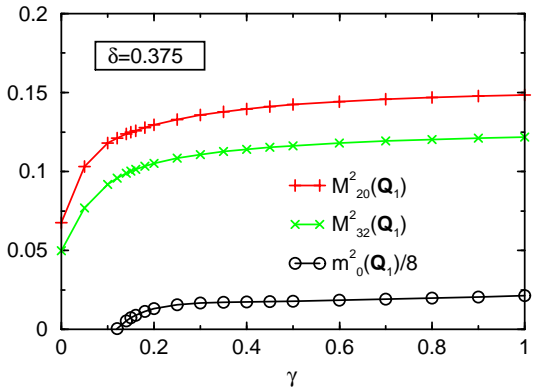}}
  \caption{(color online) $M_N^2(\vec Q_1)$ square magnetic susceptibility for $20$ and $32$ sites
    clusters and its extrapolation to thermodynamic limit $M_\infty(\vec
    Q_1)^2=m_0^2(\vec Q_1)/8$, calculated for the lowest state with momenta
    $\vec k=(0,0)$ and highest symmetry.}
\label{f:m0P375}
\end{figure}

In the case $\gamma=1$ studied in section \ref{s:CO}, we encountered a direct
transition between a collinear $(\pi,0)$ N\'eel ordered phase and  the 
Shastry-Sutherland type VBS state. In order to test this scenario here, 
we determined the static spin structure factors for different momenta in 
Fig.~\ref{f:J1mQ}. Indeed the $(\pi,0)$ components are strongest around
$\delta \sim 0.35$. In order to shed further light on the presence of magnetic
long range order we study the evolution of the collinear magnetic order as a
function of $\gamma$ for a fixed value of $\delta=0.35$. Using samples of 20 and
32 sites we obtain in Fig.~\ref{f:m0P375} a finite size scaling which shows that
the magnetic order is lost at a finite value of $\gamma\sim 0.1$, i.e. the point $\delta=0.35$,
$\gamma=0$ does not sustain magnetic long range order. 

\begin{figure*}[htbp] 
(a) \resizebox{!}{6cm}{\includegraphics{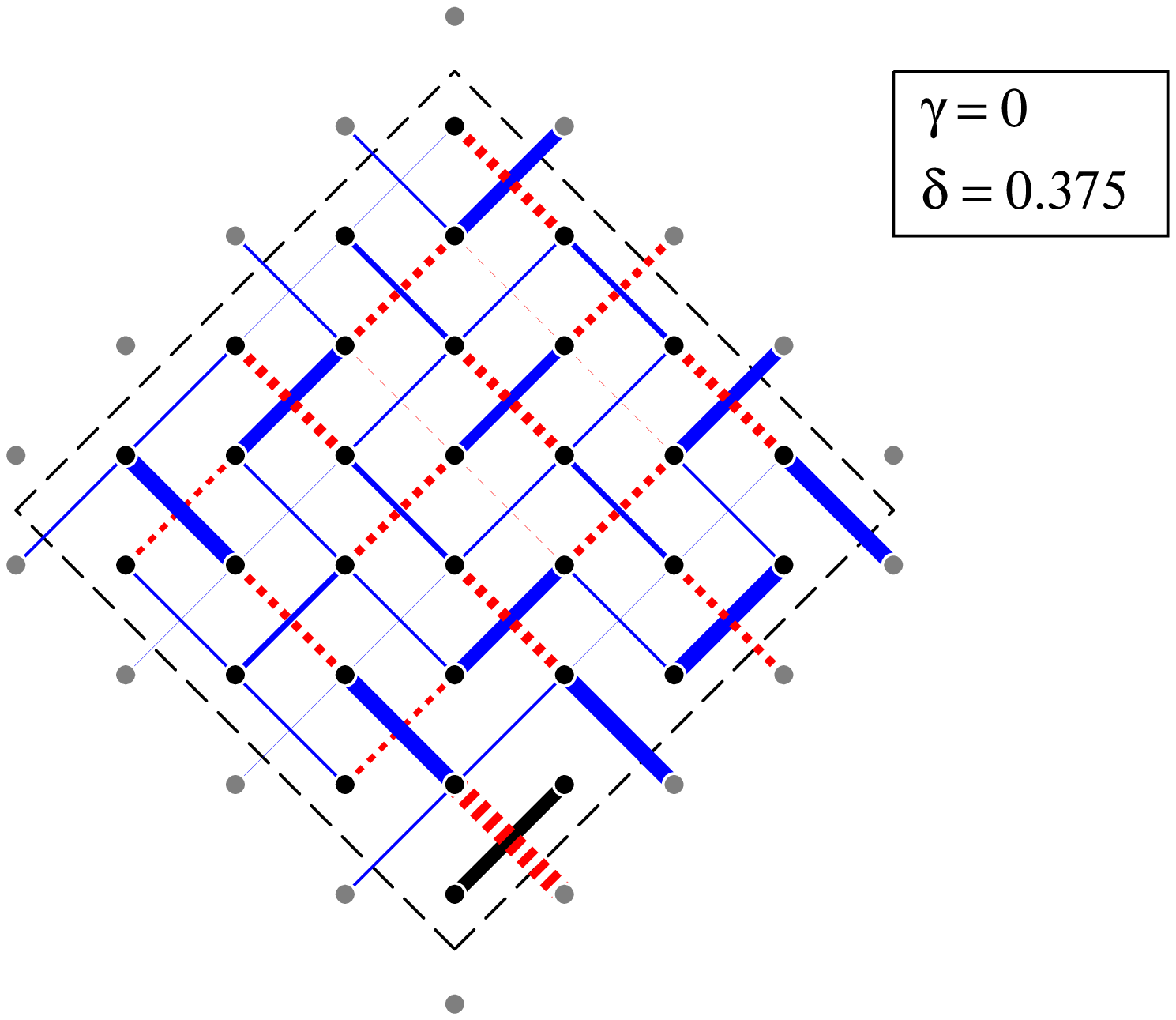}}\quad
(b)  \resizebox{!}{6cm}{\includegraphics{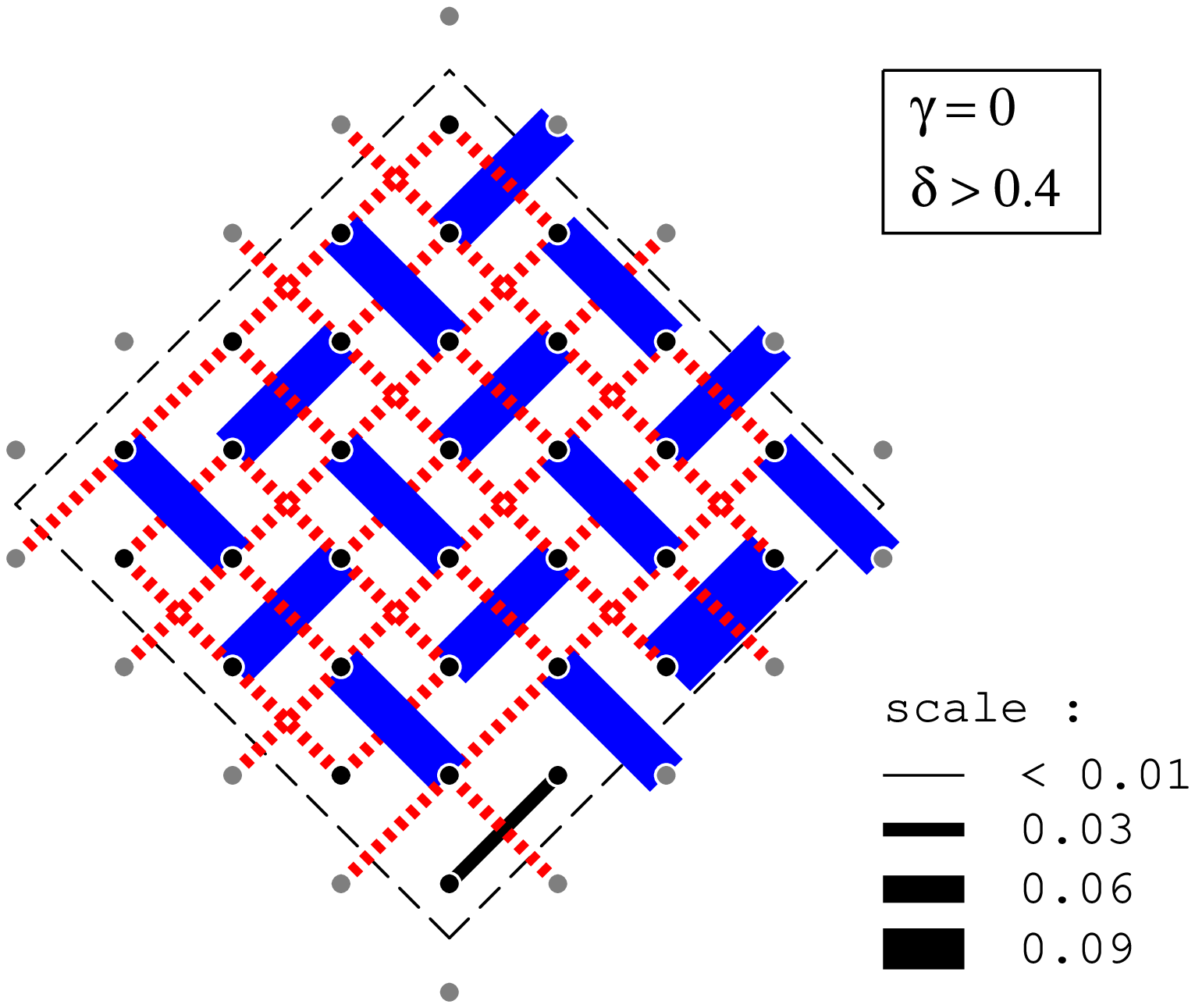}}
\caption{(color online) Representation of correlations between second neighbor
  spin dimer on the $32$ sites cluster calculated for the lowest state with
  momenta $\vec k=(0,0)$ and highest symmetry, (a) for $\gamma=0$ and
  $\delta=0.375$, (b) for $\gamma=0$ and $\delta>0.4$. Conventions are the
  same than on Fig.~\protect{\ref{f:CDim}}}
\label{f:CDSS}
\end{figure*}

Looking at real-space spin correlations, the second neighbor correlation is
much larger than all other spin correlations, so that the possibility of a dimer 
VBS state needs to be checked. We therefore also computed the real-space
dimer-dimer correlations which are presented in Fig.~\ref{f:CDSS}-(a). They
present an ordering reminiscent of the SS-VBS phase, although correlations are
smaller than in the pure phase. For the purpose of comparison the
same correlations in the pure SS-VBS phase are displayed in
Fig.~\ref{f:CDSS}-(b). We recall at this stage that in this range of
parameters, several states are very close in energy (see Fig.~ \ref{f:j1e32}),
all singlets. We also observe that they all
present the same kind of dimer-dimer correlations. From our results it is
difficult to determine if these dimer-dimer correlations are short
ranged or long ranged.
However it seems as though this phase could be best visualized by a 
condensation of singlet excitations above the SS ground states,
and not by a simple level crossing into a new ground state of completely
different character, as it seems to happen in the original Shastry-Sutherland
model \cite{SS_ED}.

{\em Phase "B" --}
Upon close inspection of Fig.~\ref{f:J1mQ} it is rather natural to suppose that there 
exists a third phase around $0.2\lesssim \delta \lesssim 0.3$ sandwiched between the phases 
"A" and "C" which displays enhanced magnetic correlations for $Q=(\pi,\pi/2)$
and $Q=(\pi/2,\pi/2)$.
Unfortunately the presence of correlations at these wavevectors 
renders the study of the system more difficult, since these momenta are not present 
on the $20$ site cluster.  It is therefore not possible to perform a finite size scaling 
of the $\vec Q$ dependent magnetic susceptibility. Nevertheless, one should note that 
there is also an enhancement of these components on the $16$ site cluster.
At that stage, it is difficult to characterize this phase "B" and to know if these 
large components correspond to a long range spin order, or to some more exotic phase.


\section{Conclusion}

We introduced a new model with frustrated interactions which provides an
interesting case of competition between antiferromagnetic orders and a valence
bond solid order.  We have shown that, for some values of the interaction
parameters, the four fold degenerate VBS ground state is exactly a direct
product of the dimer singlet wave functions. This model is thus an interesting
candidate for investigating the possibility of a newly proposed scenario of
quantum phase transition. It indeed presents, for a large range of parameters
(namely $\gamma>0.2$), a direct transition between the SS-VBS phase and a
collinear antiferromagnetic phase. Further investigations are needed to
determine if the transition is first order, of if it could correspond to the 
non-Landau-Ginsburg transition proposed by Senthil and 
coworkers~\cite{DQCP_1,DQCP_2}.

For smaller values of the second neighbor interaction
(i.e. small values of $\gamma$), the N\'eel and 
SS-VBS phases are separated by an intermediate region where
different types of correlations dominate depending on the value 
of $\delta$. It is likely that these correlations are the trace
of exotic intermediate phases, but further work is clearly needed
to fully characterize these phases and the nature of the transitions
between them.

\acknowledgments
We acknowledge the support of the Swiss National Fund and of MaNEP.
The exact diagonalization computations have been enabled by allocation of
resources on the IBM Regatta machines of CSCS Manno (Switzerland).

\appendix*
\section{}
The Hamiltonian $H_0$ (Eq.~3 of section~II) has four zero energy SS-VBS
singlet configurations forming an exact ground state. Proving (in a
mathematically rigorous way) that these four SS-VBS states are the only states
in the ground state and there exists no fifth state, is a non-trivial and hard
task. We are not going to attempt it here. Historically, for the
Majumdar-Ghosh model, which is considerably simpler as compared to $H_0$, it
was already very hard to prove the exact two-fold degeneracy of the ground
state (which was eventually shown by AKLT). However, it was much easier to
show that there are two dimerized singlet configurations which form the exact
ground state of the Majumdar-Ghosh model, and to argue that other
dimer-singlet configurations, generated by the allowed variations, will not be
the eigenstates. We will do a similar exercise for $H_0$, showing that the
four SS-VBS state are the only allowed dimer configurations in the ground
state.

The block Hamiltonian, $h_6 = P^A~P^B$, of a six-site plaquette is the basic
building block of $H_0$. Since $P^A$ and $P^B$ are the spin projectors, the
lowest energy of $h_6$ is zero. This corresponds to either $P^A$ or $P^B$ or
both becoming zero in a given spin configuration of the block. This happens
when the three $A$ sublattice spins in a six-site plaquette form a total
spin=1/2 state, or the same thing happens for $B$ sublattice spins or for
both. One way, in which this can be achieved, is by forming exactly one
singlet bond out of three $A$ or $B$ sublattice spins of a plaquette. Thus a
simple rule emerges for constructing the dimerized ground state of $H_0$. If a
dimer configuration on the full square lattice is such that, on every six-site
plaquette, there exists at least one singlet bond (dimer) between only $A$ or
only $B$ sublattice spins, then all the plaquette Hamiltonians can be
simultaneously satisfied (that is, every $h_6$ is in its ground state), and
such a configuration will be an exact zero energy ground state of $H_0$.

Now, the number of rectangular plaquettes is equal to two times the number of sites, hence
four times the number of dimers of any dimer covering of the lattice. Since a dimer
belongs at most to four rectangles, to satisfy all rectangles simultaneously, each dimer
should belong exactly to four rectangles,  and each rectangle should contain a single 
dimer. Since a dimer constructed from third neighbours belongs to only two rectangles, 
such dimers have to be rejected, and one should only use diagonal dimers.

Let us now consider one diagonal dimer. The remaining sites of the square plaquette
on which this dimer sits have to be part of a dimer. But since two dimers
cannot be on the same rectangular plaquette, the only possibility is that these 
dimers are perpendicular to the first one. This is precisely the prescription to 
construct a Shastry-Sutherland state. The freedom to chose the position and 
orientation of the first dimer leads to four different states. The exact diagonalization 
calculations on the 20 site and 32 site clusters presented in this paper support this assertion.

Note that with periodic boundary conditions of length 4, a dimer constructed
from third neighbours satisfies four rectangles, which leads to two additional ground 
states on the 16-site cluster.


\end{document}